\documentclass[prl,letterpaper,twocolumn,preprintnumbers,nofootinbib]{revtex4}
\usepackage{slashed}
\usepackage{amsmath,amssymb}
\usepackage{graphicx}
\usepackage{units}
\usepackage{xcolor}
\usepackage[hyperfootnotes=false,colorlinks,citecolor=blue]{hyperref}

\newcommand{\beq}{\begin{equation}}
\newcommand{\eeq}{\end{equation}}
\newcommand{\bea}{\begin{eqnarray}}
\newcommand{\ena}{\end{eqnarray}}

\def \epsilon {\varepsilon} 

\allowdisplaybreaks

\begin{document}

\title{
$W$~boson mass shift and muon magnetic moment in the Zee model
}

\author{\bf Talal  Ahmed  Chowdhury}
\email[E-mail: ]{talal@du.ac.bd}
\affiliation{Department of Physics, University of Dhaka, P.O. Box 1000, Dhaka, Bangladesh}
\affiliation{The Abdus Salam International Centre for Theoretical Physics, Strada Costiera 11, I-34014, Trieste, Italy}

\author{\bf Julian Heeck}
\email[E-mail: ]{heeck@virginia.edu}
\affiliation{Department of Physics, University of Virginia,
Charlottesville, Virginia 22904-4714, USA}

\author{\bf Shaikh Saad}
\email[E-mail: ]{shaikh.saad@unibas.ch}
\affiliation{Department of Physics, University of Basel, Klingelbergstrasse\ 82, CH-4056 Basel, Switzerland}

\author{\bf Anil Thapa}
\email[E-mail: ]{wtd8kz@virginia.edu}
\affiliation{Department of Physics, University of Virginia,
Charlottesville, Virginia 22904-4714, USA}


\begin{abstract}
The CDF collaboration at Fermilab has recently reported a new precision measurement of the $W$~boson mass showing a substantial $7\sigma$ deviation from the Standard Model prediction. Moreover, Fermilab has recently confirmed the longstanding tension in the $(g-2)_\mu$ measurement. We propose a unified solution to these deviations within the simplest radiative neutrino mass model: the Zee model. Our analysis establishes non-trivial links between the origin of neutrino mass, the $(g-2)_\mu$ anomaly, and the $W$~boson mass shift while being consistent with lepton flavor violation and all other experimental constraints. We find that the mass spectrum of the physical scalars must be hierarchical to be consistent with the $W$~boson mass shift; remarkably, this is also the key to resolving the $(g-2)_\mu$ tension. Furthermore, this mass splitting offers a unique same-sign dimuon signal through which our model can be tested at the LHC. 
\end{abstract}

\maketitle


\section{Introduction}
The $W$~boson mass, $M_W$, is a precisely measured quantity, and even a slight deviation from the predicted value would hint towards physics beyond the Standard Model (SM). 
In the SM electroweak (EW) fit,  $M_W$ is predicted from the following relation~\cite{Awramik:2003rn}:
\begin{align}
M_W^2|_{\rm fit}=   \frac{M_Z^2}{2}\left[1+\left(1-
    \frac{\sqrt 8 \pi \alpha_\text{EM} (1-\Delta r)}{G_F M_Z^2}\right)^{\frac12}\right],    
\end{align}
involving several precisely measured quantities: the fine structure constant $\alpha_\text{EM}$,  the Fermi constant $G_F$,  and the mass of the $Z$ boson $M_Z$.
$\Delta r=\Delta r(M_W, M_Z, m_t, \ldots)$ depends on the specific particle content of the
model and vanishes to leading order. The SM prediction for $\Delta r$ includes complete
one- and
two-loop contributions (and partial higher-order corrections), leading to
\begin{align}
M_W^{\rm SM} = (80.357 \pm 0.004) \;\textrm{GeV},
\end{align}
where the associated uncertainty originates from unknown higher-order corrections~\cite{Awramik:2003rn}. This prediction is in agreement with the most up-to-date PDG value~\cite{ParticleDataGroup:2020ssz}
\begin{align}
M_W^{\rm PDG} &= (80.379 \pm 0.012) \;\textrm{GeV},
\end{align}
at the $2\sigma$ confidence level. 

Recently, the CDF collaboration at Fermilab  has reported~\cite{CDF:2022hxs} a new precision measurement of $M_W$ using their
full $8.8$\,fb$^{-1}$ data set, yielding 
\begin{align}
M_W^\textrm{CDF-2022}= (80.4335 \pm 0.0094) \;\textrm{GeV},
\end{align}
which \emph{deviates} from the SM prediction by $7\sigma$, clearly indicating the presence of new physics (NP). 

Another important observable in the SM that has been measured in experiments with unprecedented accuracy, the muon's anomalous magnetic moment (AMM), also shows a large deviation from the SM prediction. 
This longstanding discrepancy in the muon AMM measured at BNL in 2006~\cite{Bennett:2006fi} has recently been confirmed by a new measurement performed at Fermilab~\cite{Abi:2021gix}. Combined, they correspond to a large $4.2\sigma$ disagreement with the SM prediction~\cite{Aoyama:2020ynm} that requires a NP contribution 
\begin{align}
&\Delta a_\mu^\textrm{NP} = (2.51\pm 0.59)\times 10^{-9}. \label{EXP-mu}
\end{align}

In addition to these two recent anomalies, one of the major shortcomings of the SM is its prediction of massless neutrinos. Several experiments have firmly established non-zero values of neutrino masses~\cite{ParticleDataGroup:2020ssz}, although the origin of the neutrino mass is still unknown.  

This work proposes a simultaneous explanation of all the puzzles mentioned above within the Zee model~\cite{Zee:1980ai,Zee:1985id} -- the simplest radiative neutrino mass model. We show that a consistent explanation of the CDF $W$~boson mass measurement rules out a degenerate mass spectrum of the NP scalars. It rather  dictates a mass hierarchy among the scalars, which is also crucial to resolving the longstanding tension in the $(g-2)_\mu$. Another consequence of this mass splitting is that it gives rise to a same-sign dimuon signal -- a novel feature of this model that can be uniquely probed at the LHC. By exploring the parameter space of the proposed model, we illustrate how to address these puzzles while being consistent with all current experimental data: collider bounds, charged lepton flavor violation (cLFV), EW precision observables, and low-energy measurements.

\section{Model}
\label{sec:model}
The simplest radiative neutrino mass model, the Zee model~\cite{Zee:1980ai,Zee:1985id}, utilizes two Higgs doublets $H_{1,2}$. Nonzero neutrino masses arise at the one-loop level when a singly charged scalar $\eta^+$ is introduced to this two-Higgs-doublet model framework. In the Higgs  basis (scalar potential given in Appendix-I), only one neutral Higgs acquires a nonzero vacuum expectation value, and the doublets are parameterized as,
\begin{align} 
	 H_1=\begin{pmatrix}
		G^{+}  \\
		\frac{v+\phi_1^0+iG^0}{\sqrt{2}}    
	\end{pmatrix},   
 && H_2=\begin{pmatrix}
		H^{+}_2  \\
		\frac{\phi_2^0+iA}{\sqrt{2}}     
	\end{pmatrix}, 
\end{align}
where $G^+$ and $G^0$ are the Goldstones, $H^+_2$ and $\{\phi_1^0,\phi_2^0, A \}$ are the physical scalars. The vacuum expectation value of $H_1$, $v\simeq 246$\,GeV, governs the EW symmetry breaking.

We work in the CP-conserving alignment limit~\cite{Branco:2011iw}, where the SM Higgs $\phi_1^0\approx h$ decouples from the new  CP-even Higgs ($\phi_2^0\approx H$), traditionally denoted by $\cos (\beta-\alpha)=0$, $\beta-\alpha$ being the relevant mixing angle in two-Higgs-doublet models defined as: $\sin 2(\alpha-\beta)=\frac{2 v^2 \lambda_6}{m^2_H-m^2_h}$.
Correspondingly, the masses of these states are given by,
\begin{align}
m^2_h &= \lambda_1v^2,  \  \ \ \ 
m^2_H =\mu^2_{22}+\frac{v^2}{2}(\lambda_3+\lambda_4+\lambda_5),\label{mass1}
\\
m^2_A&=m^2_H-v^2 \lambda_5, \ \  \ \ 
m^2_{H^+_2} =m^2_H-\frac{v^2}{2}(\lambda_4+\lambda_5).\label{mass2}
\end{align}
The above relations show  that these mass eigenstates can be made hierarchical, which  is the key to a simultaneous resolution of  $W$~boson mass and $(g-2)_\mu$.

For simplicity, we consider the second doublet $H_2$ to be leptophilic in nature. The leptonic Yukawa part of the Lagrangian then reads
\begin{align}
	-&\mathcal{L}_Y\supset  Y_E \bar{L} H_1 \ell_R + Y \bar{L} H_2 \ell_R + {\rm h.c.}, \label{yuk}
\end{align}
suppressing flavor indices.
In the chosen  alignment limit, the Yukawa coupling $Y_E$ associated to $H_1$ is responsible for generating masses of the charged leptons, i.e., $Y_E =\sqrt{2}\;  \mathrm{diag}(m_e,m_\mu,m_\tau)/v \equiv \sqrt{2} m_E/v$, while $Y$ determines  neutrino observables as well as the $(g-2)_\mu$.

\begin{figure}[!t]
    \centering
    \includegraphics[scale=0.25]{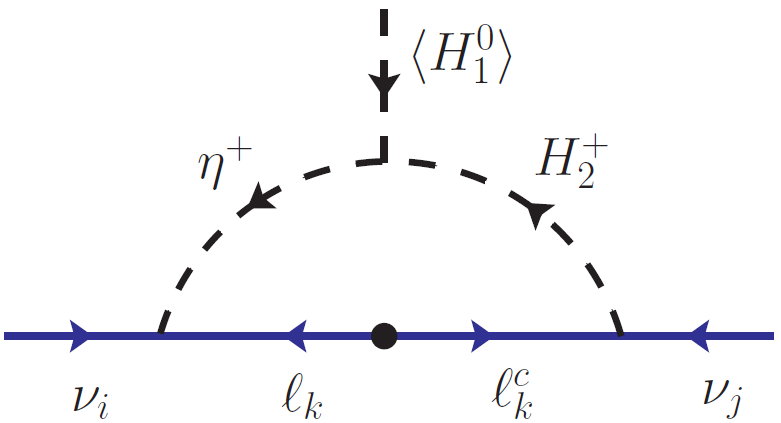}\hspace{3mm}
     \includegraphics[scale=0.25]{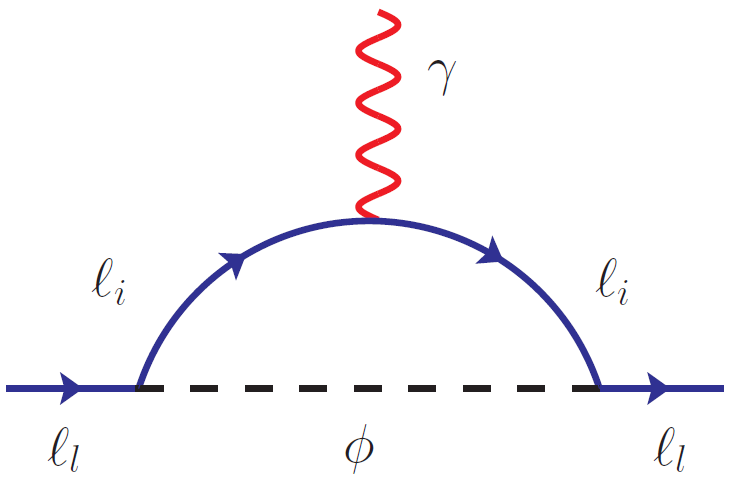}
     (a)~~~~~~~~~~~~~~~~~~~~~~~~~~~~~~~~~~~~~(b)
    \caption{(a) Radiative neutrino mass diagram in the Zee Model. (b) Corrections to $\Delta a_\ell$ in the Zee Model from the neutral scalars $\phi=H, A$. The charged scalar contribution to $\Delta a_\ell$ is obtained by replacing neutral (charged lepton) with charged Higgs (neutrino).  }
    \label{fig:loop}
\end{figure}

In addition to the Yukawa couplings given in Eq.~\eqref{yuk}, the Zee model contains one more Yukawa interaction associated with the singly charged scalar $\eta^+\sim (1,1,1)$, 
\begin{align}
-\mathcal{L}_Y\supset f_{ij}L_i \epsilon L_j \eta^+ +{\rm h.c.},\label{singly}    
\end{align}
where  $f_{ij}$ is antisymmetric in flavor indices. Eqs.~\eqref{yuk} and \eqref{singly}, together with the following cubic term in the scalar potential (rest of the terms are not relevant to our study),
\begin{align}
-V\supset \mu H_1 \epsilon H_2 \eta^- +{\rm h.c.},    
\end{align}
break the lepton number by two units and lead to a Majorana neutrino mass (left diagram in Fig.~\ref{fig:loop}) matrix~\cite{Zee:1980ai}
\begin{align}
M_\nu&=\underbrace{\frac{\sin 2\omega}{16\pi^2} \ln\left(\frac{m^2_{h^+}}{m^2_{H^+}}\right)}_{\equiv a_0}\left( f m_E Y - Y^T m_E f  \right) , \label{neutrino} 
\end{align}
where $\sin 2\omega=\sqrt{2}v\mu/(m^2_{h^+}-m^2_{H^+})$ gives the mixing angle $\omega$ between the singly charged scalars, and $h^+$ and $H^+$ represent the mass eigenstates.

\section{Correction to \texorpdfstring{$W$}{W}~boson mass}\label{sec:Wmass}

EW precision constraints can be expressed in terms of
the oblique parameters $S$, $T$, and
$U$~\cite{Peskin:1990zt,Peskin:1991sw}. In the scenario we consider, an extended Higgs sector, the NP effects enter mainly through gauge boson self-energies and $U$ is suppressed compared to $S$ and $T$. We have implemented the one-loop
corrections to these oblique parameters for our model (see  Appendix-II). In order to get the correct neutrino mass scale, the mixing angle $\omega$ between the two charged states is typically expected to be small. As a result, the flavor  eigenstate of the singly charged scalar can be approximated by its mass eigenstate (however, while computing the neutrino mass, we properly account for the mixing effects). In this limit, its effect on the oblique parameters is typically small.

\begin{figure}[!t]
    \centering
    \includegraphics[width=0.37\textwidth]{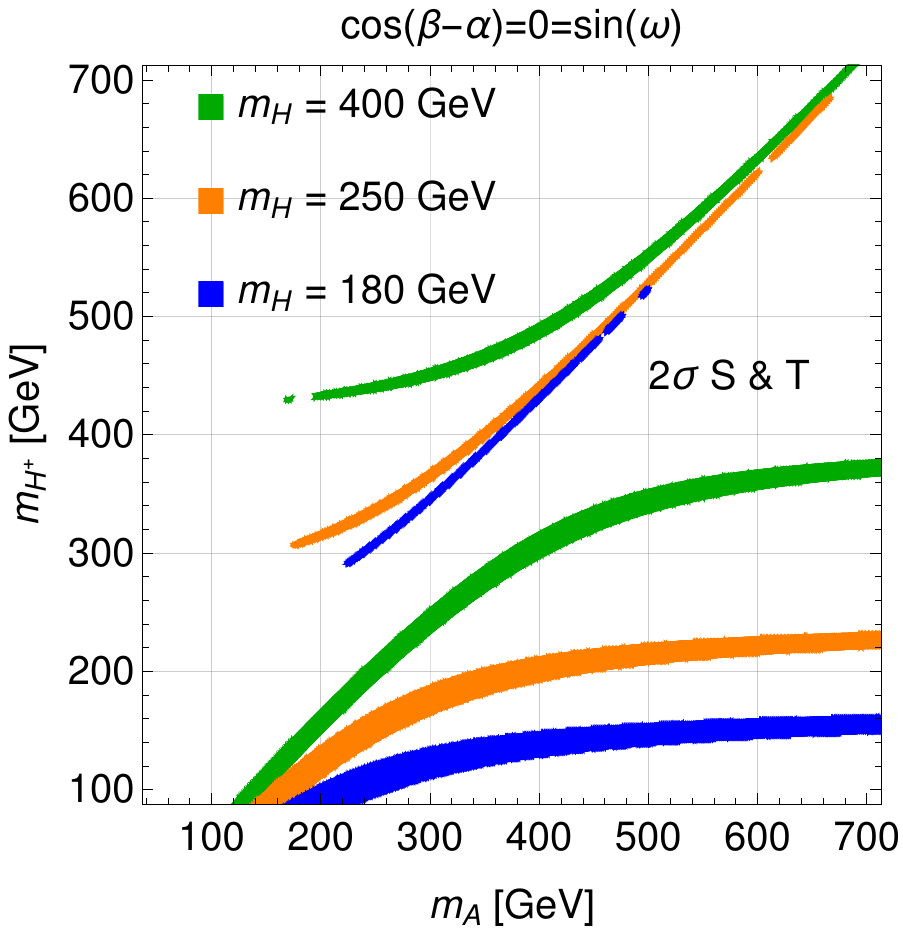}
    \caption{$2\sigma$ ranges for $S$ and $T$ from Ref.~\cite{Asadi:2022xiy} translated to masses in the aligned two-Higgs-doublet model. $\cos (\beta-\alpha)=0$ denotes the alignment limit in which $h$ and $H$ do not mix~\cite{Branco:2011iw}.}
    \label{fig:splittings}
\end{figure}

The $W$~boson mass can be calculated
as a function of these oblique parameters, given
by~\cite{Grimus:2008nb}
\begin{equation}
M_W^2 =  M^2_{W,\rm SM}
\left[
1 + \frac{\alpha_{em}\left(
c_W^2 T -\frac{1}{2} S +
\frac{c_W^2 - s_W^2}{4 s_W^2} U
\right)}{c_W^2 - s_W^2}
\right].
\label{MW-STU}
\end{equation}
Here $\theta_W$ represents the Weinberg angle. Since the EW parameters in the SM are closely related, one expects some observables in the global EW fit may suffer from new tensions once the new CDF data is taken into account. Ref.~\cite{Asadi:2022xiy} has performed a thorough global fit to relevant EW observables including the new CDF data and quoted $2\sigma$ allowed ranges for the oblique parameters that consistently explain the CDF result, which we incorporate in our analysis. We find that the mass of the charged scalar $H^+$ must be split from those of the neutral scalars $H$ and $A$ to explain the preferred $S$, $T$, $U$ ranges and therefore the $W$ boson mass measurement consistently, as illustrated in Fig.~\ref{fig:splittings}.  For a light $H^+$, we require $m_{A,H} > m_{H^+}$, while a heavy $H^+$ can also support the opposite hierarchy,  $m_{H^+} > m_{A,H}$. In order to explain the muon's AMM, we are interested in the region $m_H< m_A\lesssim m_{H^+}$. With this hierarchy, the CDF measurement together with other electroweak data imposes a lower limit on $m_H$ of around $\unit[170]{GeV}$; $m_H$ can be pushed to slightly lower values away from the alignment limit $\cos(\beta-\alpha) = 0$, but not significantly. For demonstration purpose we assumed $\sin\omega=0$ in Fig.~\ref{fig:splittings} since a tiny mixing angle $\omega$ is required to reproduce the correct neutrino mass scale.  The lower limit on $m_H$ has important consequences for the resolution of $(g-2)_\mu$ within this model.

\section{Muon's Magnetic Moment}\label{sec:g-2}
If the NP contribution of $(g-2)_\mu$ results from a single sizable coupling (for example, $Y_{e\mu}$), then the neutral component must be lighter than the charged scalar to provide the desired sign for $\Delta a_\mu$~\cite{Lindner:2016bgg}. Similarly, if a chirally enhanced contribution dominates (for example, if both $Y_{\mu\tau}$ and $Y_{\tau\mu}$ are sizable), the CP-even and the CP-odd states need to have hierarchical spectrum to avoid exact cancellations in the degenerate limit~\cite{Hou:2021sfl}. 
To satisfy all these non-trivial requirements including the new CDF measurement as well as strong experimental constraints on $H^+$, we choose $m_H< m_A < m_{H^+}$. As shown in Fig.~\ref{fig:splittings}, a  small value of $m_H$ and larger but non-degenerate values for $m_A$ and $m_{H^+}$ are fully compatible with the new CDF result. Then, choosing $m_A$ closer to $m_{H^+}$ allows for a maximum splitting between $m_H$ and $m_{H^+}$.  

In general, both the neutral and charged scalars in the model contribute to $(g-2)_\mu$; see the right diagram in Fig.~\ref{fig:loop} (which is computed in Appendix-III). However, in the aforementioned hierarchical limit, the dominant contribution arises from $H$.  Moreover, the contribution from the singly charged scalar $\eta^+$ can be ignored for heavier mass ($\sim {\rm TeV}$) and small mixing angle $\omega$. The  relevant $Y$ entries to generate the needed correction to $\Delta a_\mu$ are as follows: 
\begin{equation}
\Delta a_{\mu}\quad \Rightarrow\quad Y = \begin{pmatrix}
       . & \textcolor{red}{Y_{e\mu}} & .\\
   \textcolor{red}{Y_{\mu e}} & \textcolor{blue}{Y_{\mu\mu}} & \textcolor{teal}{Y_{\mu\tau}}\\
    . &\textcolor{teal}{Y_{\tau\mu}} & .
    \end{pmatrix}.
    \label{eq:1looptext}
\end{equation} 
In this work, we investigate three different possibilities: $\{\textcolor{red}{Y_{\mu e}, Y_{e \mu}} \}$-, $\{\textcolor{blue}{Y_{\mu \mu}} \}$-, and  $\{\textcolor{teal}{Y_{\mu \tau}, Y_{\tau \mu}} \}$-dominated scenarios. Note that only the third case is chirally enhanced due to the tau-mass flip inside the loop.
Textures differing from these three are typically disfavored from cLFV and would require significant finetuning.

\section{Low-energy Constraints/ Collider}\label{sec:constraints}

There are various low-energy constraints on the scalar mass and relevant Yukawa couplings.  The important constraints on the Yukawa couplings  $Y$ and the masses of the doublet Higgs are cLFV from radiative decay $\ell_i \to \ell_j \gamma$~\cite{Lavoura:2003xp, ParticleDataGroup:2020ssz}, trilepton decays~\cite{Cai:2017jrq}, muonium-antimuonium oscillation~\cite{Cvetic:2005gx}, as well as various experimental constraints obtained from $e^+ e^- \to \mu^+ \mu^- H$ searches at BABAR~\cite{BaBar:2016sci} and LHC~\cite{CMS:2018yxg}. Moreover, LEP~\cite{Electroweak:2003ram} puts a strong constraint with an upper limit of about \unit[30]{GeV}~\cite{Barman:2021xeq} on the scalar mass associated with $Y_{e\mu}$ to explain $\Delta a_\mu$, obtained via searches $e^+ e^- \to e^\pm \mu^\mp (H \to e^\pm \mu^\mp)$.  There are also constraints from cLFV and universality in $\ell_i  \to \ell_j \bar{\nu} \nu$ on $f_{ij}$ couplings, which are relatively relaxed compared to $Y_{ij}$ in our scenario.

\begin{figure}[tb]
    \centering
       \includegraphics[width=0.49\textwidth]{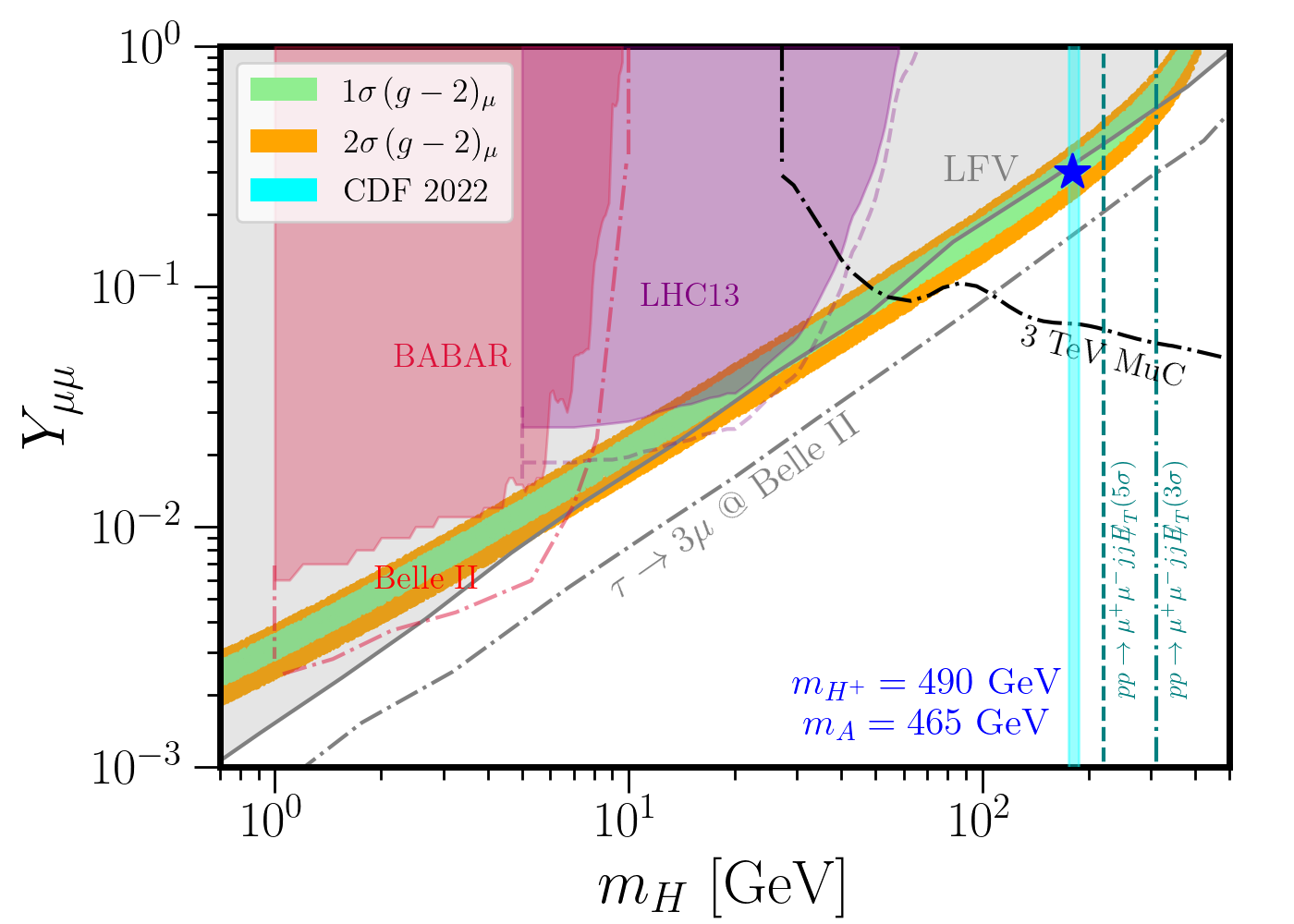}
\caption{The allowed parameter space in the Yukawa coupling and mass plane that satisfies $(g-2)_\mu$ at $1\sigma$ (green) and $2\sigma$ (orange) as well as $W$ boson mass shift (cyan). cLFV constraints are shown in grey.  The pink (purple) shaded region are excluded from $e^+e^-\to \mu^+\mu^- H$ searches at BABAR~\cite{BaBar:2016sci} and LHC~\cite{CMS:2018yxg} assuming BR$(H\to\mu\mu)=1 \, (0.5)$ with the purple dotted line  BR$(H\to\mu\mu)=1$. 
The region above the black dash-dotted line shows the projected sensitivity reach on $Y_{\mu\mu}$ from direct searches in the $\mu^{+}\mu^{-} \to \mu^{+}\mu^{-}\left(H\to \mu^{+}\mu^{-}\right)$ channel at a \unit[3]{TeV} Muon Collider assuming $\mathcal{L}=1~{\rm ab^{-1}}$~\cite{Barman:2021xeq}. The blue star denotes the benchmark fit for TX-II (the quoted values of $m_A$ and $m_{H^+}$ correspond to the benchmark fit).  The $5\sigma$ ($3\sigma$) discovery prospect  of
$pp\to \mu^\pm\mu^\pm jj \slashed{E}_T$ at the \unit[14]{TeV} LHC with integrated luminosity $\mathcal{L}=\unit[1]{ab^{-1}}$ is shown by  dashed (dash-dotted) vertical line.
}
\label{fig:g2_TX22}
\end{figure}

At a future muon collider (MuC), the coupling $Y_{\mu \mu}$ could be directly accessed via searches in the $\mu^+ \mu^- \to \mu^+ \mu^-  (H \to \mu^+ \mu^-)$ channel. The sensitivity of this channel at the projected MuC configuration of $\sqrt{s} = \unit[1]{TeV}$ with the integrated luminosity $L = \unit[1]{ab^{-1}}$~\cite{Delahaye:2019omf} is depicted in Fig.~\ref{fig:g2_TX22} (for a detailed analysis see Ref.~\cite{Barman:2021xeq}). There are no dedicated searches for the neutral scalar fields decaying into the $\tau^+ \mu^-$ sector by ATLAS and CMS. Moreover, $H^+$ can be pair-produced via the Drell--Yan process that decay into leptons $H^\pm\to \ell^\pm \nu$ ($\ell = e, \mu$). From slepton searches, LHC puts a lower limit on the charged Higgs mass $m_{H^+}\gtrsim 420$ GeV~\cite{ATLAS:2019lff} for a $100\%$ branching ratio for both $e^+e^-$ and $\mu^+\mu^+$ channels.

\section{Numerical Analysis}\label{sec:numerics}
The neutrino mass formula of Eq.~\eqref{neutrino} and the muon's AMM share the same Yukawa matrix $Y$; consequently, there are strong constraints on its elements from cLFV. Furthermore,  there is a non-trivial correlation between the $W$~boson mass shift and the $(g-2)_\mu$ that favors a specific hierarchical structure of the scalar bosons, as discussed above.      

For our numerical analysis, we consider three benchmark scenarios, each associated with a distinct texture for the $Y$ matrix, which has the following forms: 
\begin{equation}
\begin{pmatrix}
0 & \textcolor{red}{y_{e\mu}} & 0 \\
\ast & 0 & \ast \\
0 & \ast & \ast
\end{pmatrix} , \ 
 \begin{pmatrix}
\ast & 0 & \ast \\
0 & \textcolor{blue}{y_{\mu \mu}} & 0 \\
0 & \ast & 0
\end{pmatrix} , \ 
\begin{pmatrix}
0 & \ast & 0 \\
\ast & 0 & \textcolor{teal}{y_{\mu\tau}} \\
0 & \textcolor{teal}{y_{\tau \mu}} & \ast
\end{pmatrix} , 
\label{eq:FitI}
\end{equation} 
denoted as TX-I, TX-II, and TX-III, respectively. The elements shown in color are sizable and dominantly contribute to $(g-2)_\mu$; moreover, additional non-zero entries represented with ``$\ast$'' are needed to incorporate neutrino oscillation data. These textures are chosen so that we introduce a minimum number of non-zero entries to simultaneously satisfy $(g-2)_\mu$ and  observables in the neutrino sector ($\Delta m_{21}^2, \Delta m_{31}^2, \sin^2 \theta_{13}, \sin^2 \theta_{23}, \sin^2 \theta_{12}$).

\textbf{TX-I:} 
For TX-I, we find fits to neutrino observables that satisfy all cLFV. Addressing $(g-2)_\mu$ via $Y_{e\mu}$ requires $m_H< \unit[30]{GeV}$ to be consistent with bounds from LEP. This upper bound on $m_H$ along with experimental constraints on $m_{H^+}$, however,  conflicts with oblique parameters needed to satisfy the new $W$ boson mass from CDF. This rules out the TX-I scenario.    

\textbf{TX-III:} 
This texture is consistent with oblique parameters to satisfy $W$ boson mass but  inconsistent with neutrino oscillation data due to excessive cLFV. Any other combination of non-zero entries ``$\ast$'' leads to even stronger constraints from $\mu \to e \gamma$.

\textbf{TX-II:} 
The only consistent solution of $(g-2)_\mu$ and $M_W$ we obtain is for TX-II. A benchmark fit with $m_H=\unit[180]{GeV}$ for the normal hierarchy gives:  
\begin{align}
&    \theta_{12} = 32.3^\circ,\ \theta_{13}= 8.70^\circ,\ \theta_{23} = 46.6^\circ,\  \nonumber\\
  &  m_1 = 0.014\ {\rm eV},\ \delta_{CP} = 241^\circ. 
\label{eq:TXII}
\end{align}
The observables $\Delta m_{31}^2$ and $\Delta m_{21}^2$ are fitted to their respective central values \cite{Esteban:2020cvm}. Yukawa couplings associated with these fits are presented in Tab.~\ref{tab:fit}. Relevant experimental constraints and allowed parameter space to satisfy $(g-2)_\mu$ for these textures are illustrated in Fig.~\ref{fig:g2_TX22}. 
$\tau\to 3\mu$ emerges as the strongest cLFV signal and should be detectable in Belle II~\cite{Banerjee:2022xuw} (see Fig.~\ref{fig:g2_TX22}).

\begin{table*}[!t]
\centering{
    \begin{tabular}{|c|c|c|c|c|c|c|c|c|c|c|}
    \hline
          $y_{\mu\mu}$ & $y_{ \tau \mu}\, (10^{-3})$ & $y_{e\tau} \, (10^{-5})$ & $y_{ee} \, (10^{-6})$ &  $f_{e\mu}$ & $f_{e\tau}\, ( 10^{-1})$ & $f_{\mu \tau}\, ( 10^{-7})$ & $a_0\, (10^{-3})$ 
          \\[5pt]
         \hline
          $0.26 + 0.011\, i$ & $-4.86+0.23\, i$ & $2.9-0.48 \, i$ & $-5.06 - 2.9 i$ & $0.081$ & $ 2.5 + 0.23\, i$ & $4.89$ & 2.27 
          \\[5pt]
         \hline
         \hline
    \end{tabular}
    \caption{Benchmark fit for texture TX-II. }
    \label{tab:fit}
    }
\end{table*}

This model can be uniquely probed at the LHC through a novel same-sign dimuon signature~\cite{Aiko:2019mww,Jana:2020pxx}  $pp\to \mu^\pm\mu^\pm jj \slashed{E}_T$ via vector boson fusion. This process vanishes in the degenerate limit $m_H=m_A$.  The amplitude of this process is directly proportional to the mass splitting between $H$ and $A$, which is crucial in realizing $(g-2)_\mu$.
By recasting the  analysis of~\cite{Aiko:2019mww} for our benchmark scenario, the discovery prospect of $pp\to \mu^\pm\mu^\pm jj \slashed{E}_T$ signal at the \unit[14]{TeV} LHC is shown in Fig.~\ref{fig:g2_TX22}. Corresponding to this benchmark, $m_{H^+}$ up to $\sim 490\;(400)$ GeV can be discovered at $3\sigma\;(5\sigma$) CL with an integrated luminosity $\mathcal{L}=\unit[1]{ab^{-1}}$.
The Zee-model region preferred by neutrino masses, CDF, and $(g-2)_\mu$ can therefore be conclusively probed in complementary ways in the near future.

\section{Conclusions}\label{sec:conclusion}
An exciting recent development in particle physics has been the measurement of the $W$~boson mass  by the CDF Collaboration, which shows a $7\sigma$ discrepancy with the SM prediction. On top of that, the longstanding $(g-2)_\mu$ anomaly has recently been  confirmed at Fermilab. These  are clear signs for physics beyond the SM. Here,  we  showed that a simultaneous unified solution to these puzzles together with neutrino oscillation data can be achieved within the Zee model,  satisfying all current experimental constraints such as collider bounds and lepton flavor violation. In this setup, lepton flavor violation, $(g-2)_\mu$, and neutrino oscillations are inherently linked, and our detailed numerical analysis finds consistency with only a \textit{mostly muon-philic} Yukawa texture that leads to testable rates for $\tau\to 3\mu$ at Belle II.
Furthermore, our analysis shows that the mass spectrum of the physical scalars must be hierarchical to be consistent with the $W$~boson mass measured at the CDF experiment, which is also the key to solving the $(g-2)_\mu$ tension. This mass splitting offers a unique same-sign dimuon signal through which our proposed model can be tested at the LHC.

\textbf{Note added:} 
As we were completing this paper, several papers~\cite{Fan:2022dck,Zhu:2022tpr,Athron:2022qpo,Du:2022pbp,Yang:2022gvz,deBlas:2022hdk,Tang:2022pxh,Blennow:2022yfm,Zhu:2022scj,Sakurai:2022hwh,Heo:2022dey,Cheung:2022zsb,Lu:2022bgw,Strumia:2022qkt,Fan:2022yly,Cacciapaglia:2022xih,Liu:2022jdq,Lee:2022nqz,Cheng:2022jyi,Song:2022xts,Bagnaschi:2022whn,Paul:2022dds,Bahl:2022xzi,Asadi:2022xiy,DiLuzio:2022xns,Athron:2022isz,Gu:2022htv,Babu:2022pdn,Crivellin:2022fdf,Endo:2022kiw,Han:2022juu,Biekotter:2022abc,Balkin:2022glu,Kawamura:2022uft,Ghoshal:2022vzo,Perez:2022uil,Nagao:2022oin,Kanemura:2022ahw,Heckman:2022the,Ahn:2022xeq}  appeared that also discussed the impact of the new $M_W$ measurement on new physics scenarios.

\section*{Appendix-I}\label{App-01}
\textbf{Two-Higgs-doublet model potential:} 
Here we present the scalar potential of the  two-Higgs-doublet model  in the Higgs basis~\cite{Branco:2011iw}, 
\begin{align}
&V(H_1,H_2)= \mu_{1}^2H_1^{\dagger}H_1+\mu_{2}^2H_2^{\dagger}H_2
-\{\mu_{12}^2H_1^{\dagger}H_2+{\rm h.c.}\} 
\nonumber\\ & \quad
+\frac{\lambda_1}{2}(H_1^{\dagger}H_1)^2
+\frac{\lambda_2}{2}(H_2^{\dagger}H_2)^2
+\lambda_3(H_1^{\dagger}H_1)(H_2^{\dagger}H_2)
\nonumber\\ & \quad
+\lambda_4(H_1^{\dagger}H_2)(H_2^{\dagger}H_1)
+\left\{\frac{\lambda_5}{2}(H_1^{\dagger}H_2)^2+{\rm h.c.}\right\}
\nonumber\\ & \quad
+\left\{
\big[\lambda_6(H_1^{\dagger}H_1)
+\lambda_7(H_2^{\dagger}H_2)\big]
H_1^{\dagger}H_2+{\rm h.c.}\right\} .\label{pot}
\end{align}

\section*{Appendix-II}\label{App-02}
\textbf{Oblique parameters:} 

The $S$, $T$ and $U$ parameters for the Zee model in the alignment limit~\cite{Haber:2010bw} are given by~\cite{Herrero-Garcia:2017xdu}
\begin{widetext}
\begin{align}
    T &= \frac{1}{16\pi s_{w}^2 M_{W}^2}\left[c_{\omega}^2\left(F(m_{H^+}^{2},m_{H}^{2})+F(m_{H^+}^{2},m_{A}^{2})\right)+s_{\omega}^2\left(F(m_{h^+}^{2},m_{H}^{2})+F(m_{h^+}^{2},m_{A}^{2})\right)-\frac{1}{2}s_{\omega}^{2}c_{\omega}^{2}F(m_{h^+}^{2},m_{H^+}^{2})\right.\nonumber\\
    &\quad\left.-F(m_{H}^{2},m_{A}^{2})\right] ,\label{Tzee}\\
    S &= \frac{1}{\pi M_{Z}^2}\left[{\cal B}_{22}(M_{Z}^{2},m_{H}^{2},m^{2}_{A})+\frac{c_{\omega}^{2}}{2}(c_{2\omega}-3){\cal B}_{22}(M_{Z}^{2},m_{H^+}^{2},m^{2}_{H^+})-\frac{s_{\omega}^{2}}{2}(c_{2\omega}+3){\cal B}_{22}(M_{Z}^{2},m_{h^+}^{2},m^{2}_{h^+})\right.\nonumber\\
    &\quad\left.+2 s_{\omega}^{2}c_{\omega}^{2}{\cal B}_{22}(M_{Z}^{2},m_{H^+}^{2},m^{2}_{h^+})\right] ,\label{Szee}\\
    U &= -S+\frac{1}{\pi M_{W}^{2}}\left[c_{\omega}^{2}\left({\cal B}_{22}(M_{W}^{2},m_{H^+}^{2},m^{2}_{A})+{\cal B}_{22}(M_{W}^{2},m_{H^+}^{2},m^{2}_{H})\right)+s_{\omega}^{2}\left({\cal B}_{22}(M_{W}^{2},m_{h^+}^{2},m^{2}_{A})+{\cal B}_{22}(M_{W}^{2},m_{h^+}^{2},m^{2}_{H})\right)\right.\nonumber\\
    &\quad-\left. 2 c_{\omega}^{2} {\cal B}_{22}(M_{W}^{2},m_{H^+}^{2},m^{2}_{H^+})-2 s_{\omega}^{2} {\cal B}_{22}(M_{W}^{2},m_{h^+}^{2},m^{2}_{h^+})\right] ,\label{Uzee}
\end{align}
\end{widetext}
where the functions $F(x_{1},x_{2})$ and ${\cal B}_{22}(z,x_1,x_2)$ are given in~\cite{Herrero-Garcia:2017xdu, He:2001tp}, $s_\omega \equiv \sin\omega$, and $c_\omega \equiv \cos\omega$.

\section*{Appendix-III}\label{App-03}
\textbf{New physics contributions to $(g-2)_\mu$:}
Neutral scalar contribution to anomalous magnetic moment at one-loop \cite{Leveille:1977rc} as shown in Fig.~\ref{fig:loop} (b) is 
\begin{align}
    \Delta a_\ell^{H(A)}=\frac{m_{\ell}^2}{32\pi^2}\Big(\left\{|Y_{\ell i}|^2+|Y_{i\ell}|^2\right\}G[m_{H(A)},1] \nonumber \\
    \pm 2\frac{m_i}{m_\ell}\Re(Y_{\ell i}Y_{i \ell})G[m_{H(A)},0]\Big),
    \label{eq:a1N}
\end{align}
where,
\begin{equation}
    \begin{aligned}
    G(M,\epsilon)=\int_{0}^{1} \frac{x^2 - \epsilon x^3}{m_\ell^2x^2+(m_i^2-m_\ell^2)x+M^2(1-x)} \,dx.
    \end{aligned}
\end{equation}
The contribution from the charged scalar Higgs $H^+$ is obtained by replacing the neutral field with charged in Fig.~\ref{fig:loop} (b) and attaching the photon to the charged scalar, which reads
\begin{equation}
   \Delta a_\ell^{H^+}= \frac{m_\ell^2}{16\pi^2}|Y_{i\ell}|^2\int_{0}^{1} \frac{x^3-x^2}{m_\ell^2x^2+(m_{H^+}^2-m_\ell^2)x}\,dx.\label{eq:charged}
\end{equation}

\bibliographystyle{utcaps_mod}
\bibliography{BIB}
\end{document}